\documentclass[aps,prl,reprint,amsmath,amssymb,superscriptaddress,nobibnotes,showpacs,showkeys]{revtex4-1}

\usepackage{graphicx}
\usepackage{dcolumn}
\usepackage{bm}
\usepackage{amsmath}
\usepackage{amssymb,amsthm}
\usepackage{bbm}
\usepackage{epsfig,color}
\usepackage[sans]{dsfont}
\usepackage{comment}
\usepackage{hyperref}

\usepackage{color}
\usepackage{tikz}
\usepackage{tikz-3dplot}
\usetikzlibrary{patterns}
\usepackage{ifthen}
\usepackage{pgfplots}

\usepackage{graphicx}
\graphicspath{ {images/} }

\definecolor{nblue}{rgb}{0.2,0.2,0.7}
\definecolor{ngreen}{rgb}{0.2,0.6,0.2}
\definecolor{nred}{rgb}{0.7,0.2,0.2}
\definecolor{nblack}{rgb}{0,0,0}

\newcommand{\ket}[1]{|#1\rangle}
\newcommand{\ketbra}[1]{\ket{#1}\!\bra{#1}}
\newcommand{\bra}[1]{\langle#1|}

\newcommand{\tr}{\text{tr}}

\def\A{\mathcal{A}}

  \theoremstyle{definition}
  
  \theoremstyle{plain}
  
\theoremstyle{plain}
  
\theoremstyle{plain}

  \theoremstyle{plain}
  
  \theoremstyle{plain}

  \providecommand{\conjecturename}{Conjecture}
  \providecommand{\definitionname}{Definition}
  \providecommand{\lemmaname}{Lemma}
\providecommand{\corollaryname}{Corollary}
\providecommand{\theoremname}{Theorem}
\providecommand{\propositionname}{Proposition}

\def\x{\mathrm{x}}
\def\y{\mathrm{y}}

\def\i{\mathrm{id}}

\def\g{\mathrm{guess}}

\def\N{\mathcal{N}}

\def\tr{\mbox{tr}}

\def\bea{\begin{eqnarray}}
\def\eea{\end{eqnarray}}
\begin{document}


\title {Preserving Measurements for Optimal State Discrimination over Quantum Channels}

\author{Spiros Kechrimparis}
\affiliation{Department of Applied Mathematics, Hanyang University (ERICA), 55 Hanyangdaehak-ro, Ansan, Gyeonggi-do, 426-791, Republic of Korea,}
\author{Tanmay Singal}
\affiliation{Department of Applied Mathematics, Hanyang University (ERICA), 55 Hanyangdaehak-ro, Ansan, Gyeonggi-do, 426-791, Republic of Korea,}
\author{ Chahan M. Kropf}
\affiliation{ Istituto Nazionale di Fisica Nucleare, Sezione di Pavia, via Bassi 6, I-27100 Pavia, Italy, }
\affiliation{Dipartimento di Matematica e Fisica and Interdisciplinary Laboratories for Advanced Materials Physics, Universit\`{a} Cattolica del Sacro Cuore, via Musei 41, I-25121 Brescia, Italy,}
\author{ Joonwoo Bae }
\email{Corresponding author: joonwoo.bae@kaist.ac.kr}
\affiliation {School of Electrical Engineering, Korea Advanced Institute of Science and Technology (KAIST), 291 Daehak-ro Yuseong-gu, Daejeon 34141 Republic of Korea.}

\begin{abstract}
In this work, we consider optimal state discrimination for a quantum system that interacts with an environment, i.e., states evolve under a quantum channel. We show the conditions on a quantum channel and an ensemble of states such that a measurement for optimal state discrimination is preserved. In particular, we show that when an ensemble of states with equal {\it a priori} probabilities is given, an optimal measurement can be preserved over any quantum channel by applying local operations and classical communication, that is, by manipulating the quantum states before and after the channel application. Examples are provided for illustration. Our results can be readily applied to quantum communication protocols over various types of noise. 
\end{abstract}

\maketitle

Identifying a system, or distinguishing its states when there is {\it a priori} information about the possibilities, is a fundamental building block in information processing in general, and is essential to advance practical applications. In quantum information theory, optimal state discrimination \cite{ref:qsd1, ref:qsd2, ref:qsd3} is one of the key tasks in communication protocols,
e.g., the Bennett-Brassard 1984 cryptographic protocol \cite{ref:bb84}, where a measurement often aims to perform optimal discrimination for an ensemble of states.

In practical applications, however, an unwanted interaction between system and environment often exists. A quantum channel is thus naturally introduced, and it is crucial to protect the system from the intervention of noise. This is of both fundamental and practical importance for quantum information applications. In the case of quantum state discrimination, one has to deal with an ensemble of noisy states, and not the states known {\it a priori}. That is, let $S = \{ q_{\x} , \rho_{\x} \}_{\x=1}^n$ denote an ensemble of states, describing the scenario where a state $\rho_{\x}$ is generated with {\it a priori} probability $q_{\x}$. After the application of a quantum channel $\N$ on a system, the ensemble denoted by $S^{(\N)}$ is given as follows,
\bea
\N~:~ S = \{ q_{\x} , \rho_{\x} \}_{\x=1}^n ~\rightarrow~  S^{(\N)} = \{ q_{\x} , \N [  \rho_{\x} ] \}_{\x=1}^n. \nonumber
\eea
If the channel has not been verified yet, the quantum states in the ensemble $S^{(\N)}$ are not known either. 

The problem of optimal discrimination of unknown quantum states is in fact closely related to other important and long standing problems in quantum information theory. Suppose that an ensemble of quantum states can be protected by a recovery operation $\mathcal{R}$, such that $\mathcal{R}\circ \N \approx \mathrm{id}$. This may be done by preparing the ensemble in such a way that they are actually not affected by a quantum channel \cite{ref:dfs1, ref:dfs2}, or by controlling the ensemble with extra resources \cite{ref:qec1, ref:qec2, ref:qec3}. In this way, the ensemble prepared in the beginning can be retrieved, i.e., $S^{( \mathcal{R}\circ \N )} \approx S$, and therefore optimal state discrimination can be performed over the quantum channel. A recovery operation, however, may not exist for an ensemble of states in general \cite{ref:ng}. 

In order to apply the standard approach of optimal state discrimination, see e.g. \cite{ref:r1, ref:r2, ref:r3, ref:r4, ref:r5, ref:r6}, it is essential to verify a quantum channel. This, however, requires quantum process tomography, which is experimentally highly demanding. Apart from this, little is known about optimal state discrimination when a system interacts with an environment and thus evolves under a quantum channel. 

In this work, we establish the framework of optimal state discrimination over a quantum channel. Namely, we derive the conditions on a quantum channel and an ensemble of states such that a measurement for optimal state discrimination remains invariant under the channel application, i.e., an optimal measurement is preserved. Given a quantum channel, the conditions can be used to find an ensemble of states such that optimal state discrimination is preserved. Conversely, for an ensemble of states, quantum channels preserving an optimal measurement can be characterized. Remarkably, we show that when the {\it a priori} probabilities of an ensemble of states are equal, an optimal measurement can be preserved for all quantum channels in general by applying the channel twirling, which requires local operations and classical communication (LOCC) only. Examples are provided for illustration. 

\begin{figure}[]
\begin{center}
\includegraphics[width=3.4in,keepaspectratio]{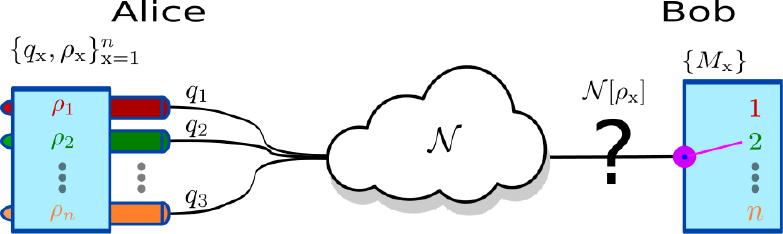}
\caption{The scenario assumes an ensemble of $n$ states $\{q_{\x}, \rho_{\x} \}_{\x=1}^n$ on Alice's side. She then chooses one of the states with probability $q_{\x}$ and sends it to Bob through a quantum channel ${\N}$ that has not been verified yet. The resulting state $\N[\rho_{\x}]$ is then unknown. Bob's measurement $\{M_{\x} \}$ aims to identify the state Alice has chosen from the ensemble.}
\label{fig}
\end{center}
\end{figure}

Let us begin by describing optimal state discrimination over a quantum channel, see Fig. (\ref{fig}). Suppose that Alice prepares a state from an ensemble $S$ and sends it to Bob through a quantum channel $\N$. Thus, it is the yet unknown ensemble $S^{(\N)}$ that is given to Bob, who then applies a measurement to learn Alice's preparation. Let $\{ M_{\x}\}$ denote a positive-operator-valued-measure (POVM) that describes Bob's measurement. The probability of obtaining outcome $\y$ given Alice's preparation $\rho_{\x}$ is given by $ \tr[M_{\y} \N [ \rho_{\x}] ]$. Then, an optimal measurement achieves the highest probability of making a correct guess on average, called the guessing probability, defined as follows,
\bea
p_{\g}^{(\N)} = \max_{ \{ M_{\x}\} } \sum_{\x=1}^n q_{\x } \tr[M_{\x} \N [ \rho_{\x}] ], \label{eq:pguessN}
\eea
where a complete measurement is performed. Note that the uncertainty about Alice's preparation given quantum states on Bob can be quantified by the conditional min-entropy, which is given in terms of the guessing probability, $H_{\min}(A|B)_{\mathrm{id}\otimes \N } = -\log p_{\g}^{(\N)}$ \cite{ref:krs}.
 
The standard problem of optimal state discrimination corresponds to the case $\N=\mathrm{id}$. In this case,  the optimization in Eq.~(\ref{eq:pguessN}) can be solved by the approach of the linear complementarity problem (LCP) in the context of convex optimization, in which the optimality conditions are analuzed \cite{ref:lcp}. The LCP finds parameters, $K$, $\{r_{\x},\sigma_{\x} \}_{\x=1}^n$, and POVM elements $\{M_{\x} \}_{\x=1}^n$ that satisfy the following optimality conditions,
\bea
K = q_{\x} \rho_{\x} + r_{\x} \sigma_{\x}~ ~\mathrm{and} ~~ \tr[M_{\x} \sigma_{\x}] = 0,~~  \x=1,\cdots, n,~~  ~~~\label{eq:opt}
\eea
where $\{r_{\x} \}_{\x=1}^n$ are non-negative numbers and $\{ \sigma_{\x}\}_{\x=1}^n $ quantum states. The parameters satisfying these conditions are automatically the solutions to the optimization problem, Eq. (\ref{eq:pguessN}). Note that the guessing probability is given by $p_{\g} = \tr[K]$, and can also be written as \cite{ref:baepra, ref:baenjp}
\bea
p_{\g}= \frac{1}{n} + r,~~\mathrm{where}~~ r = \frac{1}{n} \sum_{\x=1}^n r_{\x}, \label{eq:pg}
\eea
i.e., as the deviation from uniform by $r$.

\begin{figure}[]
\begin{center}
\includegraphics[width=3.2in,keepaspectratio]{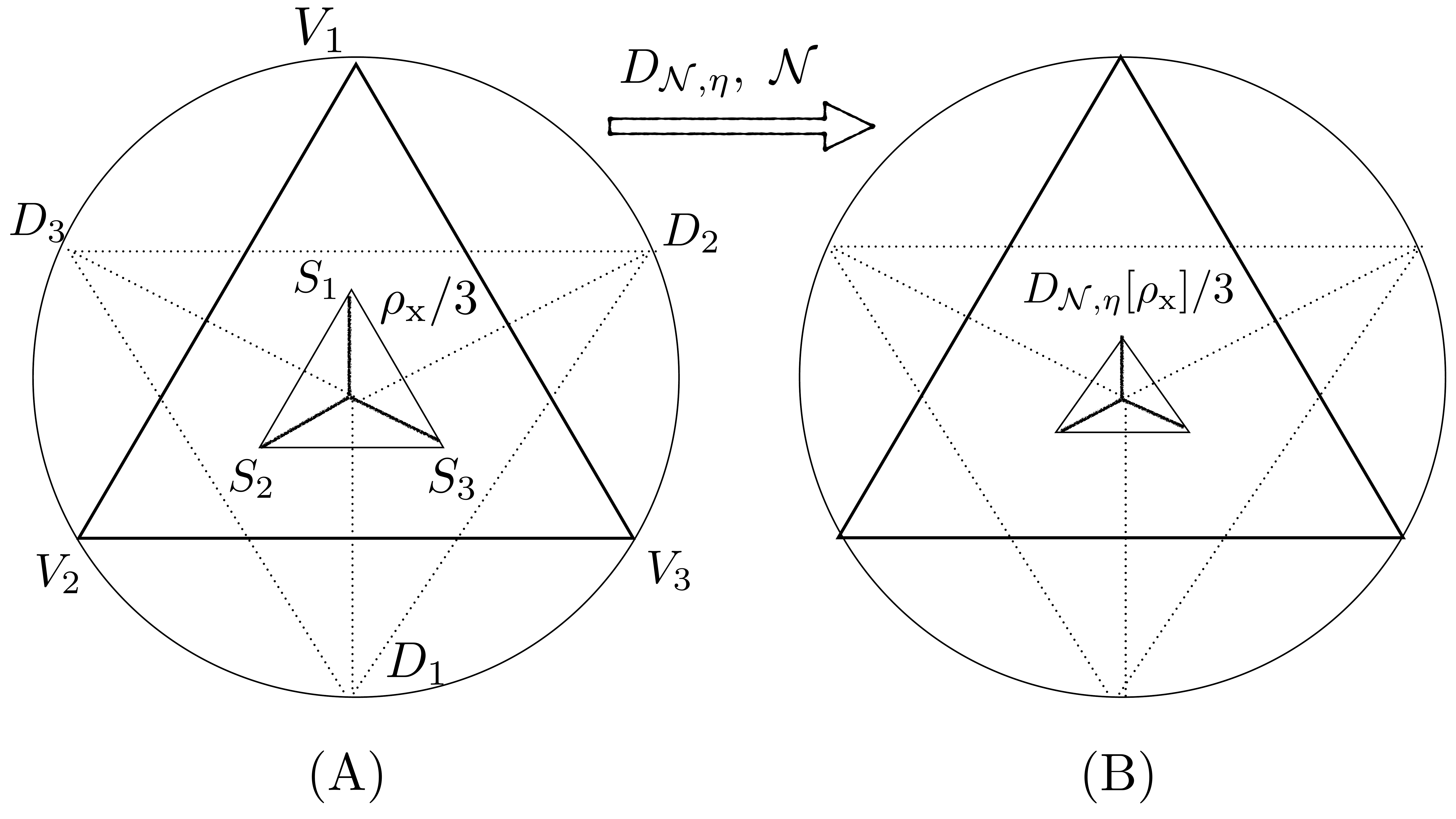}
\caption{ (A) As an illustration of the geometric approach to LCP for state discrimination, we consider an ensemble of \emph{trine} states defined as three equally distant states on the equator of the Bloch sphere, given with equal {\it a priori} probabilities $1/3$. First, one has to construct the polytope of $\{1/3 ,\rho_{\x} \}_{\x=1}^3$, which is $\triangle S_1 S_2 S_3$. Next, the latter is expanded maximally within the Bloch sphere to $\triangle V_1 V_2 V_3$. The ratio between the two polytopes corresponds to $r$, i.e., $r= OV_1 / OS_1= 1/3$, where $O$ denotes the origin. The guessing probability is therefore given by $p_{\g} = 2/3$, see Eq. (\ref{eq:pg}). By Eq.~\eqref{eq:opt2}, the states $\{\sigma_{\x} \}_{\x=1}^3$ correspond to $\{ OD_{\x} \}_{\x=1}^3$. The optimal measurement thus corresponds to $\{ OV_{\x} \}_{ \x=1}^3$, see Eq. \eqref{eq:opt}. (B) A quantum channel $\N$ may distort the geometry of the states. By the channel twirling $\mathcal{D}_{\mathcal{N},\eta}$, Eq.~\eqref{eq:tw}, the congruent geometry is reconstructed in the state space so that an optimal POVM remains the same. }
\label{figex}
\end{center}
\end{figure}

The optimal parameters in the LCP can be found by a geometric approach, as follows. Let $\mathcal{P} ( \{q_{\x},\rho_{x} \}_{\x=1}^n)$ denote the polytope in the underlying state space where each vertex is identified by $q_{\x}\rho_{\x}$, and similarly for $\mathcal{P} ( \{r_{\x},\sigma_{\x} \}_{\x=1}^n)$. Clearly, if the polytope $\mathcal{P} ( \{r_{\x},\sigma_{\x} \}_{\x=1}^n)$ is found such that the parameter $K$ in Eq. (\ref{eq:opt}) exists, it follows that an optimal measurement $\{ M_{\x}\}_{\x=1}^n$ is obtained. The conditions in Eq. (\ref{eq:opt}) can be rewritten as, 
\bea
\forall \x \neq \y, ~~q_{\x} \rho_{\x}  - q_{\y}  \rho_{\y} = r_{\y} \sigma_{\y} - r_{\x} \sigma_{\x}, \label{eq:opt2}
\eea
which gives the characterization of the polytope $\mathcal{P} ( \{r_{\x},\sigma_{x} \}_{\x=1}^n)$: the two polytopes are congruent since their edges are anti-parallel and have equal lengths. In other words, by placing the polytope $\mathcal{P} ( \{r_{\x},\sigma_{x} \}_{\x=1}^n)$ in such a way that the optimality conditions Eq. (\ref{eq:opt}) hold true for all $\x$, one can solve the problem of optimal state discrimination \cite{ref:baepra, ref:baenjp}. See Fig. \ref{figex} (A) for illustration.

We emphasize that the guessing probability, Eq. (\ref{eq:pg}), is thus obtained by referring to the geometry of the states in an ensemble, rather than pairwise relations among the states. In fact, it has been asserted that quantum distinguishability is a global property that cannot be reduced to the distinguishability of each pair of states, such as distinguishability in terms of the von Neumann entropy \cite{ref:jozsa} or the conditional min-entropy \cite{ref:baepra, ref:baenjp}. This implies that by maintaining the geometric structure of an ensemble of states, one can preserve a measurement for optimal state discrimination. \\


{\bf Definition.} A quantum channel for an ensemble of states is called  \emph {optimal measurement preserving } (OMP) if a measurement for optimal state discrimination is preserved.\\

Note that it is with respect to an ensemble of states that a quantum channel is OMP or not. We are ready to present the characterization of OMP quantum channels.  \\

{\bf Theorem.} A quantum channel $\N$ is OMP for an ensemble $S=\{q_{\x} , \rho_{\x} \}_{\x=1}^n$ if the following conditions are satisfied,
\bea
\forall \x \neq y,~   q_{\x} \N [\rho_{\x}] - q_{\y} \N [\rho_{\y}] =    \kappa ( q_{\x} \rho_{\x} -q_{\y} \rho_{y}), ~~\label{eq:pcon}
\eea
for some $\kappa \in (0,1]$. \\

{\it Proof.} Let $\{M_{\x} \}_{\x=1}^n$ denote an optimal measurement for state discrimination for an ensemble $\{q_{\x} , \rho_{\x} \}_{\x=1}^n$, and $K$ and $\{ r_{\x}, \sigma_{\x} \}_{\x=1}^n$ the corresponding optimal parameters, as in Eq. (\ref{eq:opt}). For the resulting ensemble $S^{(\N)}$, optimal discrimination can be obtained by finding parameters, denoted by $K^{(\N)}$ and $\{ r_{\x}^{( \N) }, \sigma_{\x}^{( \N)} \}_{\x=1}^n$. If we set $r_{\x}^{ (\N) } \sigma_{\x}^{( \N)} = \kappa r_{\x} \sigma_{\x} $ for all $\x$, it follows from Eq. (\ref{eq:pcon}) that,
\bea
\forall \x\neq y,~~ q_{\x} \N[\rho_{\x}] - q_{\y} \N[\rho_{\y}] = -\kappa ( r_{\x} \sigma_{\x}  - r_{\y} \sigma_{\y}  ). \nonumber
\eea
Then, we have obtained the parameters satisfying the optimality conditions: $\forall \x=1,\cdots, n$,
\bea 
K^{( \N)} = q_{\x} \N [ \rho_{\x}] + \kappa r_{\x} \sigma_{\x},~~\mathrm{and}~~\tr[M_{\x}\sigma_{\x}]=0. \nonumber
\eea
This shows that the POVM $\{ M_{\x}\}_{\x=1}^n$ is optimal for state discrimination in the ensemble $S^{(\N)}$. Note that the guessing probability is then given by,
\bea
p_{\g}^{(\N)} = \frac{1}{n}+   \kappa r,~~\mathrm{where}~~r = \frac{1}{n} \sum_{\x=1}^n r_{\x}. \label{eq:pf}
\eea
It is also shown that the guessing probability decreases with the rate $\kappa$.  $\Box$\\

The condition in Eq. (\ref{eq:pcon}) can be easily applied to find if a channel OMP is for an ensemble. For instance, consider a flip channel denoted by $\mathcal{F}$ for a pair of states such that $\mathcal{F} [\rho_{\x}] = (1 - \alpha_{\x  } ) \rho_{\x} + \alpha_{\x} \rho_{\x \oplus 1}$, where the addition $\oplus$ is in modulo $2$ and $\x=1,2$. This channel is OMP for ensembles of states with equal {\it a priori} probabilities $1/2$. This example can be further generalized to the case of additive noise, $\widetilde{\mathcal{F}}_{\gamma} = (1-\gamma)\mathcal{F} + \gamma \chi $ for some fixed state $\chi$. The channel $\widetilde{\mathcal{F}}_{\gamma}$ is also OMP.

We now restrict the consideration to ensembles with equal {\it a priori} probabilities, denoted by $S_0 = \{1/n, \rho_{\x} \}_{\x=1}^n$. It follows that $r_{\x} = r_{\y}$ for all $\x,~\y$ in the optimal condition in Eq. (\ref{eq:opt}) \cite{ref:baenjp, ref:baepra}. Then, $r=r_{\x}$ for all $\x$, see Eq. (\ref{eq:pg}). In this case, from Eq. (\ref{eq:pcon}), a depolarization channel $\Lambda_{\mu} [\rho] = (1-\mu )\rho + \mu ~\mathbb{I}/d$ for $\mu \in [0, 1]$ is OMP in general for an ensemble $S_0$. 

In fact, any quantum channel can be transformed to a depolarization map by applying the so-called channel twirling \cite{ref:dep, nielsen02, horodecki+99}. In practice, a quantum channel $\N$ can be twirled with a finite number of unitaries, called a unitary 2-design $\mathcal{U}=\{U_j\}_{j=1,\ldots,k}$ \cite{dankert+09}, as follows,
\begin{equation}
D_{\N,\eta} [\rho] = \frac{1}{k}\sum_{i=1}^{k} U_i^\dagger \N (U_i \rho U_i^\dagger) U_i =  (1-\eta) \rho+  \eta \frac{ \mathbb{I} }{d}.  \label{eq:tw}
\end{equation} 
where $D_{\N,\eta}$ denotes the map obtained after twirling the channel $\N$. The twirling protocol can thus be realized by the random application of a unitary $U_j \in \mathcal{U}$ by Alice before sending a state and then the application of its inverse $U_{j}^{\dagger}$ by Bob after receiving the state. Note that if a channel $\N$ results in a depolarization $D_{\N,\eta}$ with $\eta >1 $ by the channel twirling, one can find a unitarily equivalent one $\mathcal{N}^{'}$ for which the channel twirling in Eq. (\ref{eq:tw}) works within the range $\eta\in [0,1]$. By its inverse unitary transformation, the depolarization of the channel $\N$ with $\eta\in [0,1]$ can be obtained. 

Therefore, any channel $\N$ can be depolarized by the channel twirling such that the OMP condition in Eq. (\ref{eq:pcon}) is satisfied for an ensemble $S_0$ of arbitrary states:
\begin{equation}
\forall \x\neq \y,~~ D_{\N,\eta} [\rho_{\x}]  - D_{\N,\eta} [\rho_{\y}] = (1-\eta) (\rho_{\x}-\rho_{\y}),\nonumber
\end{equation}
with $\kappa = 1-\eta$. Note that the channel twirling does not depend on the specific choice of a unitary 2-design, see Appendix A. \\

{\bf Proposition.} Let $S_0$ denote an ensemble of quantum states with equal {\it a priori} probabilities. All quantum channels can be OMP for an ensemble $S_0$ by LOCC. \\ 

Recall that an ensemble $S_{0}^{(\N)}$ consists of unknown states when the channel $\N$ has not yet been verified. Nevertheless, an optimal measurement for an ensemble $S_0$ is immediately optimal for the ensemble $S_{0}^{(D_{\N,\eta})}$ for any channel $\N$, without the need to verify the channel. This makes it possible to perform optimal state discrimination over a quantum channel even without identification of the channel or the resulting ensemble. 

This has many applications on the practical side. On the one hand, an experimental setup prepared in a laboratory for optimal state discrimination for a given ensemble can still be used to perform an optimal measurement in some other environment, such as an open-space realization. On the other hand, if the ensemble $S_0$ is unknown {\it a priori} and an optimal measurement is obtained for the ensemble $S_{0}^{(D_{\N,\eta})}$ for some channel $\N$, then the measurement is immediately also optimal for the unknown ensemble $S_0$. This resolves the problem of optimal extraction of classical information from quantum states that are exploited to deliver classical messages over a quantum channel. As an illustration of Proposition, we demonstrate the latter case in what follows.

Let us consider the ensemble $S_0$ of a pair of qubit states $\rho_j = (\mathbb{I}+\boldsymbol{r}_{j}\cdot\boldsymbol{\sigma})/2$ for $j=1,2$ with $\boldsymbol{r}_1 = (1,1,0)/\sqrt{2}$ and $\boldsymbol{r}_2 = (-3,3\sqrt{3},0)/8$, respectively. The polytope of the states corresponds to the line $(\rho_1 - \rho_2)/2$. We consider a \emph{bit-phase flip} channel, $ \N_p [\rho]=(1-p) \rho + p Y \rho Y^\dagger$, with $p=0.45$, that is not OMP for the ensemble $S_0$. By twirling the channel $\N_p$, the resulting channel is given by $D_{\N_p, \eta_0}$, with $\eta_0 = 4p/3$, which is OMP.

\begin{figure}[]
\begin{center}
\includegraphics[width=3.4in,keepaspectratio]{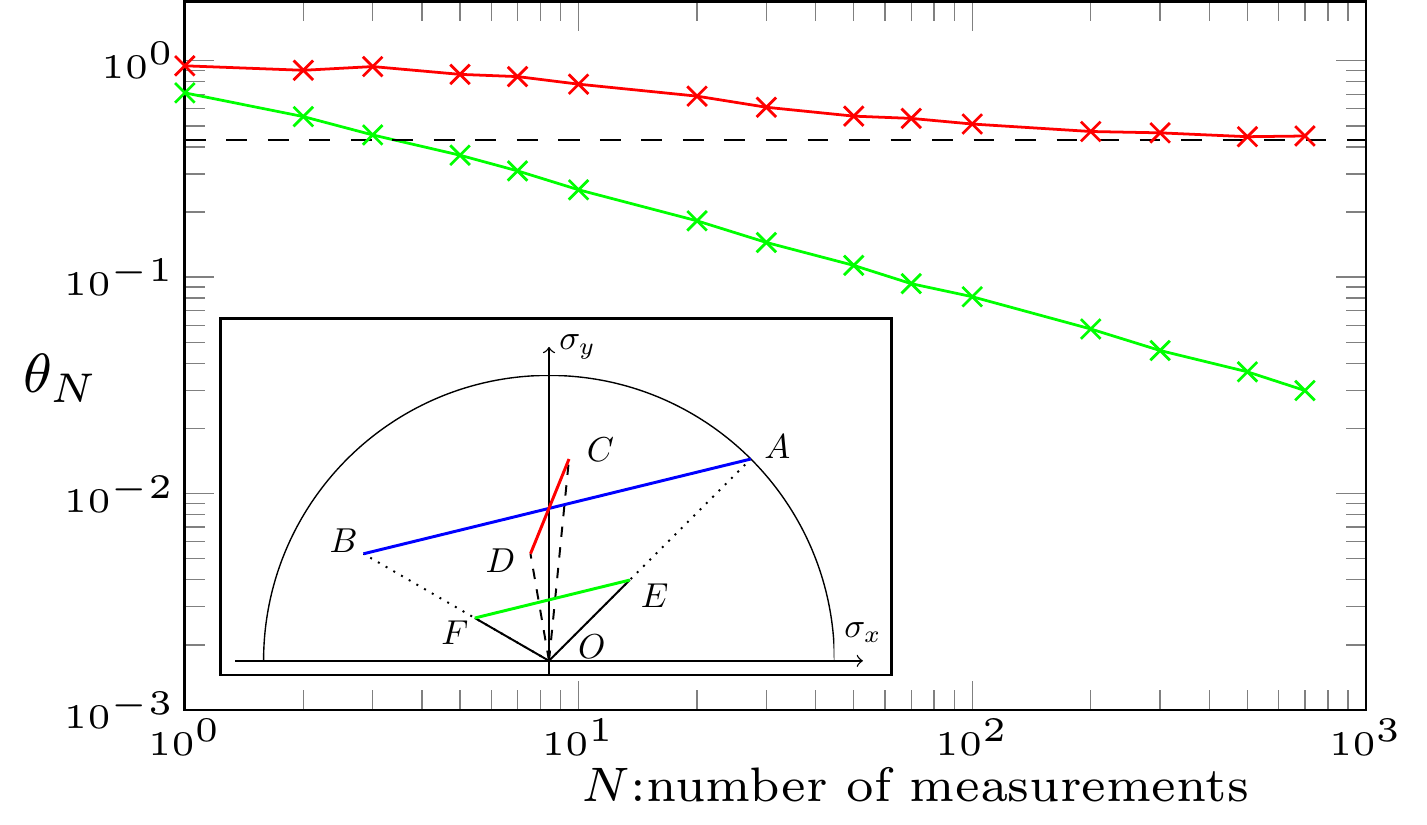}
\caption{ The parameter $\theta_N$ from reconstruction of the state in Eq. (\ref{eq:ef}) shows convergence to the optimal measurement, depending on whether the channel twirling is applied or not. (Inset) The states $\rho_1$ and $\rho_2$, denoted by $OA$ and $OB$, are sent through the non-OMP quantum channel, $\N_p$. The resulting states correspond to $OC$ and $OD$, respectively. By the channel twirling, two states whose difference corresponds to $EF$ that is parallel to $AB$ are obtained, i.e., the optimal measurement is preserved. (Main panel) Without the channel twirling, the optimal measurement is not preserved as $\theta_N$ does not converge to zero (red line). By the channel twirling, the angle converges to zero as $1/\sqrt{N}$ (green line). }
\label{gch}
\end{center}
\label{fig: optimal POVM} 
\end{figure}

In the inset of Fig.~\ref{gch}, $OA$ and $OB$ describe states $\rho_1$ and $\rho_2$, respectively, where $O$ denotes the origin. The optimal measurement lies on the diameter parallel to the difference $AB$. After the states are sent through the channel, resulting states $\N_p[\rho_1]$ and $\N_p[\rho_2]$ correspond to $OC$ and $OD$, respectively. The difference $CD$ is not parallel to $AB$, and thus the optimal measurement of the resulting states is not equal to that of the original states. By the channel twirling, we have two states $D_{\N_p,\eta_0} [\rho_1]$ and $D_{\N_p,\eta_0} [\rho_2]$ that correspond to $OE$ and $OF$, respectively. The difference $EF$ is parallel to $AB$, and thus the corresponding optimal measurement is equal to the optimal discrimination between the states $\rho_1$ and $\rho_2$ given in the beginning.
 
In order to estimate $CD$ and $EF$, we devised the following operator as the object to reconstruct from the counting statistics,
\begin{equation}
\rho_{\Lambda} =\frac{1}{2}( \mathbb{I} + \Lambda [ \rho_1]- \Lambda[ \rho_2]). \label{eq:ef}
\end{equation}
with $\Lambda = D_{\N_p , \eta_0 }$ with the channel twirling, and $\Lambda = \N_p$ without it, respectively. 

For the channel twirling, we here employ the well-known Clifford group containing $24$ elements, see Appendix A. For the reconstruction of the operator in Eq. (\ref{eq:ef}), we perform the maximum-likelihood estimation from the simulated counting statistics of the measurement in the basis $\sigma_x$, $\sigma_y$ and $\sigma_z$. In Appendix B, the specifics pertaining to the maximum-likelihood reconstruction are explained. In Fig. \ref{gch}, the performance is demonstrated with $N$ measurements on each basis $\sigma_x$, $\sigma_y$, and $\sigma_z$. The figure of merit, $\theta_N$, is taken as the angle between $AB$ and the estimated $CD$ and $EF$, respectively, averaged over $1000$ realizations. Without the channel twirling, the angle converges to a finite number, which shows that the optimal measurement is not preserved. By the channel twirling, the angle converges to zero, meaning that the optimal measurement is preserved.

It is worth comparing the guessing probabilities in the three cases: i) the ideal case $\N=\mathrm{id}$ , ii) a quantum channel $\N$, and iii) a twirled channel $D_{\N,\eta}$. For an ensemble $S_0$ of $n$ states, it is straightforward from Eq. (\ref{eq:pg}), see also Eq. (\ref{eq:pf}), that
\bea
p_{\g}^{(\i)} = \frac{1}{n} + r ~ \geq ~ p_{\g}^{(  D_{\N,\eta})} = \frac{1}{n} +\eta r. \label{eq:rel} 
\eea
Hence, it is clear that the guessing probability does not increase under a channel action. In the example shown above, we observe that $p_\mathrm{guess}^{(\mathrm{id})} = 0.77$, and $p_{\g}^{(  D_{\N_p, \eta_0 })} = 0.61 > p_{\g}^{( \N_p)} = 0.53$, i.e., the guessing probability can increase by the channel twirling for some channels.

In conclusion, we have considered optimal state discrimination when states are sent through a quantum channel, and established the framework for the preservation of an optimal measurement. We have characterized the OMP quantum channels over which a measurement of optimal state discrimination remains invariant. In technical terms, the characterization provided in Theorem, see Eq. (\ref{eq:pcon}), shows that if the geometric structure of an ensemble of states is preserved, so is any optimal measurement for the ensemble. It is worth mentioning that the conditions for the preservation are obtained in terms of pairwise relations of quantum states. Finally, we have shown that for ensembles with equal {\it a priori} probabilities, any quantum channel is, after channel twirling, OMP, which is feasible with LOCC. This has been demonstrated with two-state discrimination.  
 
Our results are not only fundamental for understanding the relations between states, channels, and measurements, but also useful for practical applications such as communication protocols. For instance, from our results it follows that if an interaction between system and environment describes a non-OMP channel, a measurement does not have to be prepared again for optimal state discrimination by the channel twirling. In future investigations, it would be interesting to find how the information-preserving structure for quantum states \cite{ref:ng2} is related to OMP channels. In the former, the guessing probability of pairwise states is preserved, and in the latter, which we have shown here, an optimal measurement for the guessing probability is preserved. It would also be interesting to investigate the capacities of OMP channels. We may envisage that our results can be used to improve quantum protocols for reliable communication. 

 \section*{Acknowledgement}
This work is supported by and Institute for Information and communications Technology Promotion (IITP) grant funded by the Korea government (MSIP) (No. R0190-15-2028, PSQKD) and National Research Foundation of Korea (NRF2017R1E1A1A03069961), ITRC Program(IITP-2018-2018-0-01402), and Iniziativa Specifica INFN-DynSysMath.

\section*{Appendix A: Twirling and unitary 2-designs \label{Appendix: twirling}}

In this appendix we list some known facts about twirling and unitary 2-designs. The first is that twirling over the Haar measure produces a depolarizing channel \cite{nielsen02,horodecki+99}, that is,
\begin{equation}
\int_{U(d)} U^\dagger \mathcal{N} (U\rho U^\dagger) U \, dU = (1-\eta) \rho+  \eta \frac{ \mathbb{I} }{d} \,, \label{eq:integral twirling}
\end{equation}
where the parameter $\eta$ by depends on the channel $\mathcal{N}$ only. To be precise, 
\begin{align*}
	1- \eta = \frac{\sum_k |\textrm{tr}(A_k)|^2-1}{d^2-1},
\end{align*}
where $A_k$ are the Kraus operator of the channel $\mathcal{N}$, i.e., $\mathcal{N}(\rho) = \sum_kA_k^\dagger\rho\A_K$.
The second is the defining property of a unitary 2-design $\mathcal{U}=\{U_j\}_{j=1,\ldots,k}$ with $k$ elements  in total:
\begin{equation}
\frac{1}{k}\sum_{i=1}^{k} U_i^\dagger \mathcal{N}(U_i \rho U_i^\dagger) U_i= \int_{U(d)} U^\dagger \mathcal{N} (U\rho U^\dagger) U \, dU \,.
\end{equation}
Thus, instead of having to randomly draw unitaries from $U(d)$ to perform the integral in Eq.\@ \eqref{eq:integral twirling}, it suffices to twirl over the finite elements of the unitary 2-design, i.e. 
\begin{equation}
\mathcal{D}_{\mathcal{N},\eta}[\rho]\equiv\frac{1}{k}\sum_{i=1}^{k} U_i^\dagger \mathcal{E}(U_i \rho U_i^\dagger) U_i = (1-\eta) \rho+  \eta \frac{ \mathbb{I} }{d} \,, \label{eq:twirling}
\end{equation}
By definition, the result of twirling does not depend on the specific choice of a unitary 2-design. The protocol described in the main text becomes practically more efficient if a \emph{minimal unitary 2-design} is chosen, that is, one with the least number of elements. In Ref. \cite{gross+07} a lower bound on the number of elements was proven, found to be $d^4-2d^2+2$, where $d$ is the dimension of the Hilbert space, while a $d^4-d^2$ was conjectured to be a tight bound. 

Let us now describe two choices of a unitary 2-design in the case of $d=2$. One such choice could be the widely used Clifford group which has 24 elements (up to phase factors). Let $\mathcal{C}$ denote the Clifford group; then, its elements are explicitly 
\bea
\mathcal{C} &=&  \{\mathbb{I},X,Y,Z,H,S,XH,XS,YH,YS,ZH,ZS, HS, \nonumber \\
&&  SH,XHS,XSH,YHS,YSH,ZHS,ZSH, HSH, \nonumber \\
&&  XHSH,YHSH,ZHSH  \}. \label{eq:Cliff}
\eea
where $\mathbb{I}$ denotes the identity matrix, $X,Y,Z$ the Pauli matrices, $H$ the Hadamard matrix, while S denotes the phase matrix with $\varphi=\pi/2$. 

Another possible choice, the number of elements of which saturates the conjectured tight lower bound for a minimal unitary 2-design in $d=2$, is a subgroup of the tetrahedral group of rotations only, which consists of 12 elements \cite{gross+07,bengtsson+06}. Let $\mathcal{R}$ denote the aforementioned subgroup of the tetrahedral group. Then its elements are given by
\begin{equation}
\mathcal{R} = \{\mathbb{I} , e^{-i \pi \frac{\boldsymbol{e} \cdot \boldsymbol{\sigma} }{2}}, e^{-i \frac{\pi}{27} \boldsymbol{r} \cdot \boldsymbol{\sigma}},e^{-i \frac{2\pi}{27} \boldsymbol{r} \cdot \boldsymbol{\sigma}} \} \,,
\end{equation}
where $\boldsymbol{e}$ is shorthand for the three elements obtained by the substitution with the three vectors $\boldsymbol{e}_1= (1,0,0) \,, \boldsymbol{e}_2= (0,1,0)$ and $\boldsymbol{e}_3= (0,0,1)$. Similarly $\boldsymbol{r}$ is shorthand for the four elements $\boldsymbol{r}_1=\boldsymbol{e}_1 +\boldsymbol{e}_2 +\boldsymbol{e}_3$, $\boldsymbol{r}_2=\boldsymbol{e}_1 -\boldsymbol{e}_2 -\boldsymbol{e}_3$, $\boldsymbol{r}_3=-\boldsymbol{e}_1 -\boldsymbol{e}_2 +\boldsymbol{e}_3$ and $\boldsymbol{r}_4=-\boldsymbol{e}_1 -\boldsymbol{e}_2 +\boldsymbol{e}_3$.

\section*{Appendix B: Maximum-likelihood reconstruction \label{Appendix: reconstruction}}

In the main text we considered the reconstruction of the state \eqref{eq:ef} from the simulated counting statistics of the measurement in the Pauli basis. 
%
There are various methods that one can employ when it comes to the problem of quantum state reconstruction, two popular approaches being \emph{maximum likelihood} and \emph{Bayesian} tomography. In our simulations (c.f. Fig~\ref{gch}), we employed the \emph{maximum-likelihood} state reconstruction, which we will now briefly review. 
	
	Assume that we have $n$ copies of a quantum state $\rho$ on which we perform measurements to obtain information about the state. The experimentally observed data is encoded as a sequence of events $E: \{e_1\,,\ldots\,,e_n\}$, with $e_j=(\alpha,k)$ indicating that outcome $k$ was observed when measurement $\alpha$ was applied, represented theoretically by the operator $\Pi_{\alpha,k}$. 
	
	In the example we considered in the main text, we have $3N\cdot 24$ copies of each of the states $\rho_1$ and $\rho_2$. We simulated events for the measurements in the Pauli basis $\{\sigma_x,\sigma_y,\sigma_z\}$ with outcomes $\pm x,\pm y,\pm z$. In other words, $\Pi_{\alpha,k} = \ketbra{\pm\alpha}$ with $\alpha = x,y,z$ and $k=+,-$. Without twirling, we simulated the measurements for each of the two states $24N$ times. In the case of twirling, we simulate events for the same measurements for each of the two states and for each twirling element in Eq.~\eqref{eq:Cliff} $N$ times. In both cases we thus obtain $n=2\cdot3\cdot 24 N$ events $e_j$ in total. These events can then be used to reconstruct the state Eq.~\eqref{eq:ef} in the following way.
			
	With $n_{\alpha,k}$ denoting the number of events with outcome $k$ when measurement $\alpha$ was applied, and with $f_{\alpha,k}=n_{\alpha,k}/n$ the relative frequencies, then the \emph{likelihood} functional of the observed data $E$ given the quantum state $\rho$ is defined as
	\begin{equation}
	\mathcal{L}(E|\rho)\equiv\prod_{\alpha,k} \left(p_{\alpha,k}\right)^{n_{\alpha,k}}=\left(\prod_{\alpha,k} \left(\tr\left(\rho \Pi_{\alpha,k}\right)\right)^{f_{\alpha,k}}\right)^n \,, \label{eq:likelihood functional}
	\end{equation}
	where $p_{\alpha,k}=\tr\left(\rho \Pi_{\alpha,k}\right)$ is the theoretical probability of obtaining outcome $k$ after measurement $\alpha$ was performed. In maximum likelihood estimation (MLE), one is maximizing the likelihood functional, Eq.~\eqref{eq:likelihood functional}, over the set of density matrices; that is, one is looking for the quantum state that is \emph{most likely} to have produced the obtained data. 
	
	In practice, one can also take the logarithm of the likelihood functional and restate the problem as the following optimization problem:
	\begin{equation*}
	\begin{aligned}
	& \underset{\rho}{\text{maximize}}
	& & \log\mathcal{L}(E|\rho)=\sum_{\alpha,k}f_{\alpha,k}\, \log p_{\alpha,k} \\
	& \text{subject to}
	& &  \rho\geq 0 \text{ and } \tr(\rho)=1.
	\end{aligned}
	\end{equation*}
	The convexity of the logarithm guarantees that it attains a maximum value.
	
	The latter optimization problem can thus be encoded in the definition of the following \emph{log-likelihood} functional
	\begin{align*}
	\mathcal{F}(\rho)=\sum_{\alpha,k} f_\alpha^k \log P_\alpha ^k - \mu (\tr(\rho)-1)\,,
	\end{align*}
	where $\mu$ is a Langrange multiplier that has been added to ensure that the matrix $\rho$ is indeed of trace one. Note that the positivity and Hermiticity of $\rho$ will be implicitly enforced by Eq.~\eqref{eq:extremal eq3}.
	
	Setting the variation of the functional equal to zero,
	\begin{equation*}
	\frac{\delta\mathcal{F}(\rho)}{\delta \rho}=0\,,
	\end{equation*}
	leads to the condition
	\begin{equation}
	R-\mu \mathbb{I} =0 \,, \label{eq:extremal eq1}
	\end{equation}
	that a state needs to obey in order to be an extremum of $\mathcal{F}(\rho)$.
	The  matrix $R$ is defined as
	\begin{equation*}
	R\equiv \sum_{\alpha,k} \frac{f_{\alpha,k}}{p_{\alpha,k}}\Pi_{\alpha,k} \,,
	\end{equation*}
	and it implicitly depends on the state $\rho$ though the theoretical probabilities $p_{\alpha,k}$. We choose to rewrite the extremal equation \eqref{eq:extremal eq1} in the form
	\begin{equation}
	R \rho R= \mu^2 \rho \label{eq:extremal eq2} \,,
	\end{equation}
	in order to guarantee the Hermiticity of the density matrix $\rho$. Using the constraint $\tr(\rho)=1$, we obtain the value of the Lagrange multiplier, $\mu=\sqrt{\tr(R\rho R)}$.
	Putting it all together, the extremal equation becomes
	\begin{align}
	\frac{R \rho R}{\tr(R \rho R)}= \rho \label{eq:extremal eq3} \,.
	\end{align}
	Due to the fact that the matrix $R$ depends implicitly on the state $\rho$, Eq.\@ \eqref{eq:extremal eq3} is of course non-linear. Although an analytic solution is not possible in general, an approximate solution can be obtained by iterating Eq.\@ \eqref{eq:extremal eq3} according to the prescription
	\begin{align}
	\rho^{(n+1)} =\frac{R^{(n)} \rho^{(n)} R^{(n)}}{\tr(R^{(n)} \rho^{(n)} R^{(n)})}\label{eq:iterative extremal eq} \, \,, \quad n=0,1,2\ldots,m\,,
	\end{align}
	where $m$ is the maximum number of iterations in the numerical routine. The initial choice must be a proper density matrix, i.e., $\rho^0 = {\rho^0}^\dagger$, and $\rho^0\geq 0$.
	A possible choice is the maximally mixed one, i.e. $\rho^{(0)}= \mathbb{I} /d$.

\end{document}